\documentclass[aps]{revtex4}
\usepackage{epsf}
\begin{document}
%
\title{Polymers in long-range-correlated disorder}
\author{V.~Blavats'ka}
\email[]{viktoria@icmp.lviv.ua}
\affiliation{Institute for Condensed
Matter Physics of the National Academy of Sciences of Ukraine,
79011 Lviv, Ukraine}
\author{C. von Ferber}
 \email[]{ferber@thphy.uni-duesseldorf.de}
\affiliation{Institut f\"ur Theoretische Physik II,
Heinrich-Heine-Universit\"at D\"usseldorf, 40225 D\"usseldorf, Germany}
\author{Yu.~Holovatch}
\email[]{hol@icmp.lviv.ua}
\affiliation{Institute for Condensed
Matter Physics of the National Academy of Sciences of Ukraine,
79011 Lviv, Ukraine}
\affiliation{Ivan Franko National University of Lviv, 79005 Lviv, Ukraine}
\begin{abstract}
  We study the scaling properties of polymers in a $d$--dimensional
  medium with quenched defects that have power law correlations
  $\sim r^{-a}$ for large separations $r$. This type of disorder is
  known to be relevant for magnetic phase transitions. We find strong
  evidence that this is true also for the polymer case. Applying the
  field-theoretical renormalization group approach we perform calculations
  both in a double expansion in $\varepsilon=4-d$ and $\delta=4-a$ up to
  the 1-loop order and secondly in a fixed dimension ($d=3$) approach up
  to the 2-loop approximation for different fixed values of the
  correlation parameter, $2\leq a \leq 3$. In the latter case the
  numerical results need appropriate resummation. We find that  the
  asymptotic behavior of self-avoiding walks in three dimensions and
  long-range-correlated disorder is governed by a set of
  separate exponents. In particular, we give estimates for the $\nu$ and
  $\gamma$ exponents as well as for the correction-to-scaling exponent
  $\omega$. The latter exponent is also calculated for the
  general $m$-vector model with $m=1,2,3$.
\end{abstract}
\pacs{64.60.Fr,61.41.+e,64.60.Ak,11.10.Gh}
\date{\today}
\maketitle

\section{INTRODUCTION}\label{I}
The influence of structural disorder on the critical behavior of
various kinds of condensed matter remains one of the central problems
in physics. In this paper, we are interested in the scaling laws that
govern the behavior of polymers in disordered media when the defects
are correlated or belong to some porous or sponge-like structure. Our
main question of interest will be: does a small amount of correlated
quenched structural defects in the medium induce changes to the
universal properties of a polymer macromolecule?

It is well established that the universal scaling properties of long
flexible polymer chains in a good solvent are perfectly described
within a model of self-avoiding walks (SAWs) on a {\em regular}
lattice \cite{desCloizeaux90}.  The limit of SAWs with an infinite
number of steps may be mapped to a formal $m\rightarrow 0$ limit of
the $m$-vector model at its critical point \cite{deGennes79}. In
particular, for the average square end-to-end distance $R_e$ and the
number of configurations $Z_N$ of a SAW with $N$ steps on a regular
lattice one finds in the asymptotic limit $N\to\infty$:
\begin{equation}\label{nu}
 \langle R_e^2 \rangle \sim N^{2\nu},\mbox{\hspace{3em}}
Z_N \sim z^{N} N^{\gamma-1}
\end{equation}
where $\nu$ and $\gamma$ are the universal correlation length and
susceptibility exponents for the $m=0$ model that only depend on the
space dimensionality $d$ and $z$ is a non-universal fugacity.
For $d=3$ the exponents read \cite{Guida}
$\nu=0.5882\pm 0.0011$ and $\gamma=1.1596\pm0.0020$;
 whereas for $d=2$ exact values $\nu=3/4$ and $\gamma=43/32$ are
known \cite{d2}.

The problem of SAWs on {\em randomly diluted} lattices, which may
serve as a model of linear polymers in a porous medium,
has been the subject of intensive discussion
\cite{Chakrabarti,Kim,Harris83,Meir,Grassberger93,Barat,randomsaws,%
Ordemann,Blavats'ka01}.
A recent review on SAW statistics on random lattices is given in
ref.~\cite{Barat}.  The numerical results for these systems
available from Monte-Carlo simulations, exact enumeration and
analytical treatment also cover the non-universal properties.
Nonetheless, even apart from the numerical values of scaling exponents
the question if a given form of disorder affects the scaling behavior
has not been settled in general.

A frequently studied type of random lattice is the lattice that is
diluted to the percolation threshold \cite{Barat}.  Here, one is
interested in the behavior of a SAW on the percolation cluster.
Scaling laws (\ref{nu}) hold with exponents that differ from
their counter parts on a regular lattice at $d=2$ and $d=3$
\cite{Meir,Grassberger93}. Apparently, this results from the fact that
the percolation cluster itself is characterized by a fractal dimension
which differs from $d$, the Euclidean one. Moreover, the scaling of
the averaged moments of $Z_N$ (\ref{nu}) on the backbone of a
percolation cluster possesses multifractal behavior \cite{Ordemann}.
In our study, however, we address another type of disorder, when the
lattice is well above the percolation threshold. In this case the
dimension of the support does not change and it is not clear {\em a
priori} whether the SAW asymptotic exponents (\ref{nu}) will be
influenced.

We approach this question for the case of long-range-correlated disorder
using the connection between the scaling properties of polymers and magnets.
So let us first turn our attention to the magnetic problem. While it is
intuitively clear that strong disorder destroys the magnetic
ordering, a much more subtle question is what happens at weak dilution by a
non-magnetic component, i.e. well above the percolation threshold
(weak disorder) \cite{Folk00}. It has been argued \cite{Harris} that the
presence of  a non-correlated (or short-range-correlated)
quenched disorder has a nontrivial effect on the critical
behavior  of magnetic systems, only if the specific heat critical
exponent $\alpha$  of the pure magnet is positive. This statement is
often called the Harris criterion. However, one should be careful in
applying this ``naively" to the SAW problem. Indeed, although the
critical exponent $\alpha$ of a SAW on the $d=3$ dimensional pure lattice
is positive \cite{Guida} ($\alpha(d=3)=0.235\pm0.003$), a weak quenched
short-range-correlated disorder does not alter the SAW critical
exponents. This statement has been proven by Harris \cite{Harris83}
and confirmed later by  renormalization group results \cite{Kim}.

Note, that in the works mentioned above only uncorrelated quenched
defects were investigated. In 1980-ies the model of a disordered
$d$-dimensional system with so-called ``extended'' structural defects
\cite{Cardy,Dorogovtsev} was developed. These defects are considered
as quenched and correlated in a subspace of $\varepsilon_d$
dimensions, and randomly distributed in the remaining
$d-\varepsilon_d$ dimensions. This model may be applied for small
densities of defects. The integer values of $\varepsilon_d$ have a
direct physical interpretation: $\varepsilon_d=0$ corresponds to
short-range-correlated point-like defects, and the cases
$\varepsilon_d=1,2$ are related, respectively, to lines and planes of
impurities. To give an interpretation to non-integer values of
$\varepsilon_d$, one may consider patterns of extended defects like
aggregation clusters, and treat $\varepsilon_d$ as the fractal
dimension of these clusters \cite{Yamazaki88}.
In this interpretation the defect patterns are fractal,
while the support of the system is the complement of this fractal
and will in general not be fractal itself.
In ref. \cite{Cardy} the critical behavior of
$O(m)$ symmetric magnets with extended defects with parallel
orientation was investigated evaluating the renormalization group
equations by a double $\varepsilon=4-d,\varepsilon_d$ expansion. It
was found that the scaling is affected by these kinds of defects and
the critical exponents were calculated in this scheme. The static and
dynamic critical properties of $m$-component cubic-anisotropic systems
with extended-defects were studied in the refs. \cite{Yamazaki} using
a double
$\varepsilon,\varepsilon'=\varepsilon+\varepsilon_d$-expansion again
finding a change of critical behavior when $\varepsilon_d$ is
increased.

In further work \cite{Weinrib} attention concentrated on disordered
systems with ``random-temperature'' disorder, arising from a small
density of impurities that cause random variations in the local
transition temperature $T_c(\vec{x})$. The fluctuations in
$T_c(\vec{x})$ are characterized by a correlation function, that falls
off according to a power law: $\sim x^{-a}$ at large distances $x$. It
was shown, that in the presence of long-range-correlated disorder the
Harris criterion is modified: for $a<d$ the disorder is relevant, if
the correlation length critical exponent of the pure system obeys
$\nu<2/a$. An $m$-vector model of this type was evaluated using a
renormalization group expansion in the parameters
$\varepsilon=4-d,\delta=4-a$ up to the linear approximation.  An
additional renormalization group fixed point corresponding to the
long-range correlated disorder was found.  In the following we will
denote this as the `LR' fixed point.  The correlation-length exponent
was evaluated in this linear approximation as $\nu=2/a$ and it was
argued, that this scaling relation is exact and also holds in higher
order approximation. However, this result was questioned recently in
refs. \cite{Prudnikov}, where the static and dynamic properties of 3d
systems with long-range-correlated disorder were studied in a
renormalization group approach using a 2-loop approximation.  There is
an essential discrepancy between the latter results and those found
from the $\varepsilon,\delta$-expansion. Nevertheless it was
qualitatively confirmed by both approaches that long-range-correlated
disorder leads to a new universality class for these magnetic systems.
Note, that the variable $a$ is a global parameter: together with the
space dimension $d$ and the number of components $m$ of the order
parameter it fixes the universal values of the critical exponents.

While the influence of long-range-correlated disorder on the magnetic
phase transition has been the subject of considerable interest, the
effect of long-range-correlated disorder on the scaling properties of
polymers remains unclear and is generally not considered as settled.
Here, we address the question of the asymptotic behavior of polymers
in long-range-correlated disorder with algebraically decaying
correlations \cite{Weinrib}. While the linear approximation of the
double $\varepsilon,\delta$-expansion indicates qualitatively
the existence of the LR fixed point for polymers, it leads to unphysical
quantitative results \cite{Blavats'ka01}.  For this reason we present
here an analysis of the 2-loop approximation using the fixed
$a,d$-technique that leads to physically meaningful results for the
scaling behavior of polymers in the LR regime.

Our paper is organized as follows: in the following section \ref{II}
we present the model, in section \ref{III} the renormalization
procedure is discussed and we reproduce the results of the
$\varepsilon,\delta$-expansion. In section \ref{IV} we apply
resummation techniques to analyse the renormalization group functions
in the two-loop approximation and find that the asymptotic behavior of
self-avoiding walks in a three-dimensional medium with
long-range-correlated disorder is governed by a new set of exponents.
For the exponents we present quantitative estimates. Section \ref{V}
concludes our study.  Some additional information about the properties
of magnetic phase transitions in systems with long-range-correlated
quenched disorder is presented in the appendix.


\section{THE MODEL}\label{II}
To study the universal properties of polymers in porous media with
long-range-correlated quenched structural defects, we turn our
attention to the investigation of the appropriate $m$-vector model in
the polymer limit. We consider the model of an $m$-vector magnet, that
is described by the following Hamiltonian:
\begin{equation}
{\cal H}=\int
d^dx\left[\frac{1}{2}\left((\mu_0+\delta\mu_0(x))\vec{\phi}^2+(\vec{\nabla}
\vec{\phi})^2\right)+\frac{u_0}{4!}(\vec{\phi}^2)^2\right],
\end{equation}
here, $\vec{\phi}$ is an $m$-component field:
$\vec{\phi}=\{ \phi^{1}\cdots\phi^{m}\}$,
$\mu_0$ and $u_0$ are the bare mass and the coupling of the undiluted
magnetic model, $\delta\mu_0(x)$ represents the quenched random-temperature
disorder, with:
$$
 \langle\langle\delta\mu_0(x)\rangle\rangle=0,
$$
$$
 \frac{1}{8}\langle\langle\delta\mu_0(x)\delta\mu_0(y)\rangle\rangle
=g(|x-y|),
$$
where $\langle\langle\cdots\rangle\rangle$ denotes an average over
spatially homogeneous and isotropic quenched disorder. The form of the
pair correlation function $g(r)$ is chosen to fall off with distance
according to a power law \cite{Weinrib}:
\begin{equation}
g(r) \sim r^{-a} \label{1}
\end{equation}
for large $r$, where $a$ is a constant.

We consider quenched disorder and
average the free energy over different configurations of
the disorder. To this end we apply the replica method and construct an
effective Hamiltonian for the $m$-vector model with
long-range-correlated disorder \cite{Weinrib}:
\begin{eqnarray}
 {\cal H}_{\rm eff}&=&\sum_{\alpha=1}^{n}\int{\rm d}^dx \Big[
\frac{1}{2}\left(\mu_0\vec{\phi}_{\alpha}^2+
(\vec{\nabla}\vec{\phi}_{\alpha})^2\right) +
\frac{u_0}{4!} (\vec{\phi}_{\alpha}^2)^2\Big]
{}-\sum_{\alpha,\beta=1}^{n}
\int{\rm d}^dx{\rm d}^dy g(|x-y|)
\vec{\phi}_{\alpha}^2(x)\vec{\phi}_{\beta}^2(y).
\label{4}
\end{eqnarray}
Here, the replica interaction vertex $g(r)$ is the correlation
function given in Eq.~(\ref{1}), Greek indices denote replicas and the replica
limit $n\to 0$ is implied.

For small $k$ the Fourier-transform $\tilde g(k)$ of $g(r)$ reads:
\begin{equation}
\tilde g(k)\sim v_0+w_0|k|^{a-d}.
\label{2}
\end{equation}
Note, that in the case of random uncorrelated point-like defects the
site-occupation correlation function formally reads:
$ g(|x-y|)\sim \delta(|x-y|),$
and its Fourier transform obeys:
\begin{equation}
\tilde g(k)\sim v_0.
\label{3}
\end{equation}
Comparing Eqs.~(\ref{2}) and (\ref{3}), it is obvious that the case
$g(r)\sim r^{-d}$ corresponds to random uncorrelated point-like
disorder.  Moreover, different integer values of $a$ correspond to
uncorrelated extended impurities of random orientations.  So, the
correlation function in Eq.~(\ref{1}) with $a=d-1$ describes
straight lines of impurities of random orientation whereas random
planes of impurities correspond \cite{Cardy} to $a=d-2$.
In terms of the fractal interpretation given in the introduction,
the general case corresponds to SAWs on the {\em complement} of a fractal
with dimension $\epsilon_d=d-a$.

Writing Eq.~(\ref{4}) in momentum space and taking Eq.~(\ref{2})
into account, one obtains an effective Hamiltonian with three bare
couplings $u_0,v_0,w_0$.  For $a>d$ the $w_0$-term is irrelevant in
the renormalization group sense and one obtains the effective
Hamiltonian of the quenched diluted (uncorrelated) $m$-vector
model \cite{Grinstein} with two couplings $u_0,v_0$. For $a<d$ we
have, in addition to the momentum-independent couplings, the momentum
dependent one $w_0k^{a-d} $.  Note that $\tilde g(k)$ must be
positive being the Fourier image of the correlation function.
This implies $w_0\geq0$ for small $k$. Also the coupling $u_0$
must be positive, otherwise the pure system would undergo a first order
phase transition.

The critical behavior of the model in Eq.~(\ref{4}) with $m\geq 1$
has been investigated \cite{Weinrib,Prudnikov,Elka} using the
renormalization group approach \cite{rgbooks}.  We are interested in the
polymer
limit $m\rightarrow 0$ of this model interpreting it as a
model for polymers in a disordered medium.  Note, that this limit is
not trivial.  For the case $u_0\neq 0, v_0\neq 0, w_0=0$ the ``naive''
RG analysis leads to controversial results about the absence of a
stable fixed point and thus to the absence of the second order phase
transition \cite{Chakrabarti}.  As noticed by Kim \cite{Kim}, once the
limit $m,n\to 0$ has been taken, the $u_0$ and $v_0$ terms are of the
same symmetry, and an effective Hamiltonian with one coupling of
$O(mn=0)$ symmetry results.  This leads to the conclusion that  weak
quenched uncorrelated disorder is irrelevant for polymers as long as
$v_0<u_0$.

Our present analysis takes these symmetry properties into account.
In the case of the Hamiltonian with a term
for long-range-correlated disorder Eq.~(\ref{4}) we pass
to an effective Hamiltonian \cite{Blavats'ka01} with only two
couplings $U_0= u_0-v_0$ and $w_0$ (in what follows below we will keep
the notation $u_0$ for this new coupling $U_0$). In discrete momentum
space this effective Hamiltonian reads:
\begin{eqnarray} \label{h}
{\cal H}_{\rm eff}&=&\sum_{k}\sum_{\alpha}^n \frac{1}{2}
(\mu_0+k^2)(\vec{\phi}_k^{\alpha})^2
+ \frac{u_0}{4!}\sum_{\alpha}^n\sum_{k_1 k_2 k_3k_4}
\delta(k_1+k_2+k_3+k_4)
\left( \vec{\phi}_{k_1}^{\alpha}\vec{\phi}_{k_2}^{\alpha} \right)
\left( \vec{\phi}_{k_3}^{\alpha}\vec{\phi}_{k_4}^{\alpha} \right)-
\\ \nonumber
&& \frac{w_0}{4!}\sum_{\alpha \beta}^n
\sum_{kk_1k_2k_3k_4}|k|^{a-d}\delta(k_1+k_2+k)\delta(k_3+k_4-k)
\left(\vec{\phi}_{k_1}^{\alpha} \vec{\phi}_{k_2}^{\alpha}\right)
\left(\vec{\phi}_{k_3}^{\beta}\vec{\phi}_{k_4}^{\beta}\right), \quad m,n
\rightarrow 0.
\end{eqnarray}
Here, the $\delta(k)$ represent products of Kronecker symbols
and the notation $\left( \vec{\phi}\vec{\phi} \right)$
implies a scalar product.
Note that the $w_0$-term itroduces interactions between the replicas and
contains the power of an internal momentum. Again, it may be shown
that for $a=d$  in the limit $m,n
\rightarrow 0$ the $u_0$ and $w_0$ terms are of the same symmetry
and one is left with an $O(mn=0)$-vector model with only one coupling
$(u_0-w_0)$.


\section{THE RENORMALIZATION}\label{III}
In order to extract the critical behavior of the model, we use the
field-theoretical renormalization group
(RG) method.  We choose the massive field theory
scheme with renormalization at non-zero mass and zero external momenta
\cite{Parisi} that leads to Callan-Symanzik equations for the
renormalized one-particle irreducible vertex functions
$\Gamma^{(M)}_R$. In our case the renormalization conditions \cite{Prudnikov}
are written both in fixed $d$ and $a$. The renormalized
mass $\mu$ and renormalized couplings $u, w$ are defined by:
\begin{eqnarray}
\mu^2&=& \Gamma_R^{(2)}(k,\mu^2,u,w)|_{k=0}, \nonumber\\
\mu^{4-d} u &=&\Gamma_{R,u}^{(4)}(\{k\},\mu^2,u,w)|_{k=0},  \nonumber\\
\mu^{4-a} w &=&\Gamma_{R,w}^{(4)}(\{k\},\mu^2,u,w)|_{k=0}.  \nonumber
\end{eqnarray}
Here, $\Gamma^{(4)}_{R,u}$ and $\Gamma^{(4)}_{R,w}$ are the
contributions to the four-point vertex function $\Gamma^{(4)}_R$ that
correspond to $u$- and $w$-term symmetry, respectively.
Asymptotically close to the critical point the $M$-point renormalized
vertex functions obey the homogeneous Callan-Symanzik  equation
\cite{rgbooks}:
\begin{equation}
\left\{\mu\frac{\partial}{\partial
\mu}+\sum_i\beta_{v_i}(\{v_j\})\frac{\partial}{\partial
v_i}-\frac{M}{2}\gamma_{\phi}(\{v_j\})\right\}
\Gamma^{(M)}_R(\{k\},\mu^2,\{v_j\})=0,
\end{equation}
here, $v_i=u,w$.  The change of the couplings $u,w$ under
renormalization defines a flow in parametric space that is governed by
the corresponding $\beta$-functions $\beta_{u}(u,w)$,
$\beta_{w}(u,w)$.  The fixed points $u^{\ast},w^{\ast }$ of this flow
are given by the solutions of the system of equations: $
\beta_u(u^{\ast},w^{\ast})=0,\, \beta_w(u^{\ast},w^{\ast})=0.$ The
stable fixed point is defined as the fixed point where the stability
matrix:
\begin{equation}
B_{ij}=\frac{\partial \beta_{v_i}}{\partial v_{j}},
\label{5}
\end{equation}
possesses eigenvalues $\omega_i$ with positive real parts.  The
accessible stable fixed point corresponds to the critical point of the
system. The fixed point is accessible if it can be reached along flow
lines starting from allowed initial values $u_0,w_0\geq 0$.  At the
fixed point we define the correlation length and pair correlation
function critical exponents $\nu$ and $\eta$ by
\begin{eqnarray}\label{d1}
\nu^{-1}&=&2-\gamma_{\phi}(u^{\ast},w^{\ast})-
\gamma_{\phi^2}(u^{\ast},w^{\ast}), \\
\label{d2}
\eta&=&\gamma_{\phi}(u^{\ast},w^{\ast}),
\end{eqnarray}
where  $\gamma_{\phi^2}$ is the exponent that corresponds to the
two-point vertex function $M=2$ with a $\phi^2$ insertion. Other critical
exponents may be obtained from familiar scaling laws. For example,
for the susceptibility exponent $\gamma$ one has:
\begin{equation} \label{d3}
\gamma= \nu(2-\eta).
\end{equation}
According to the RG prescriptions given above, the RG functions are
obtained in form of a series in the renormalized couplings. In the
one-loop approximation the result reads \cite{Blavats'ka01}:
  \begin{equation} \beta_u = - \varepsilon
\left[ u- \frac{4}{3}\,u^2 I_1 \right ] -\delta 2uw \left[
I_2+\frac{1}{3} I_4 \right] + (2 \delta- \varepsilon)
\frac{2}{3}\,w^2 I_3 , \label{A} \end{equation}
 \begin{equation} \label{A1}
\beta_w = -\delta \left [w + \frac{2}{3}\,w^2 I_2  \right ]
+\varepsilon  \frac{2}{3} \left[wu I_1- w^2 I_4 \right] ,
\end{equation}
 \begin{equation} \gamma_{\phi^2}= \varepsilon
\frac{u}{3}\, I_1 -\delta \,\frac{w}{3}\,I_2 , \phantom{55}
\gamma_{\phi}=\delta\, \frac{w}{3} I_4.  \label{B}
 \end{equation}
Here, $I_i$ are one-loop integrals that depend on the space dimension
$d$ and the parameter $a$:
\begin{equation} \label{int}
I_1= \int\frac{ {\rm d}\vec{q}}{(q^2 + 1)^2 },\,\phantom{5} I_2=
\int\frac{ {\rm d}\vec{q}\, q^{a-d}}{(q^2 + 1)^2 },\,\phantom{5} I_3=
\int\frac{ {\rm d}\vec{q},\, q^{2(a-d)}}{(q^2 + 1)^2
 }\,\phantom{5} I_4=\frac{\partial }{\partial k^2}\left [ \int\frac{
{\rm d}\vec{q}\, q^{a-d}}{[q+k]^2 + 1 }\right ]_{k^2=0}.
\end{equation}
Note that contrary to the usual $\phi^4$ theory the $\gamma_{\phi}$
function in Eq.~(\ref{B}) is nonzero already in one-loop order. This is
due to the $k$-dependence of the integral $I_4$ in Eq.~(\ref{int}).

\begin{figure} [!htb]
\epsffile{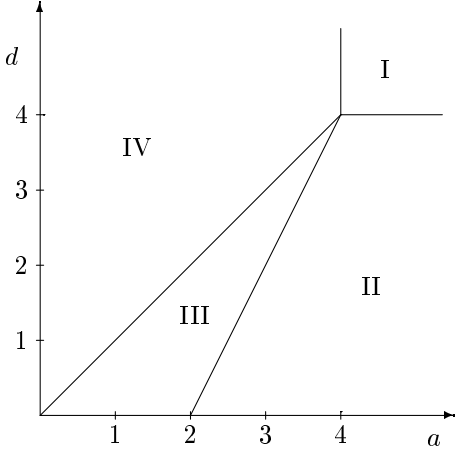}
\caption{\label{fig1} The critical behavior of a polymer in a medium with
long-range-correlated disorder in different regions of the $d,a$-plane as
predicted by the first order $\varepsilon,\delta$--expansion. Region I
corresponds to the Gaussian random walk behavior, in the region II scaling
behavior is the same as in the medium without disorder, in  region III
the ``long-range'' fixed point LR is stable and the scaling laws for polymers
are altered, in region IV no accessible stable fixed points appear; this may
be interpreted as the collapse of the chain.
 } \end{figure}

There are two ways to proceed in order to obtain the qualitative
characteristics of the critical behavior of the model.
One can consider the polynomials in Eqs. (\ref{A}), (\ref{A1}) for fixed
$a,d$ and look for
the solution of the fixed point equations. It is easy to check that these
one-loop equations do not have any stable accessible fixed points for $d<4$.
The other scheme
to evaluate these equations is a double expansion in
$\varepsilon=4-d$ and $\delta=4-a$ as proposed by Weinrib and Halperin
\cite{Weinrib}. Formerly \cite{Blavats'ka01}, we exploited this
up to the linear approximation. For completeness, we here note those
results. Substituting the loop integrals in Eqs.~(\ref{A})-(\ref{B})
by their expansion in $\varepsilon=4-d$ and $\delta=4-a$ , one
obtains the 3 fixed points given in thea Table \ref{tab1}. We may draw
the following conclusions from these first order results:
Three distinct accessible fixed points are found to be stable
in different regions of the $a,d$-plane. The Gaussian (G) fixed point, the
pure (P) SAW fixed point and the long-range (LR) disorder SAW fixed point.
The corresponding  regions in the $a,d$-plane are marked by I, II and III in
fig.~\ref{fig1}. In the region IV no stable fixed point is accessible.

For the correlation length critical exponent of the SAW, one finds
distinct values
$\nu_{\rm pure}$ for the pure fixed point and $\nu_{\rm LR}$ for the
long-range fixed point.  Taking into account
that the accessible values of the couplings are $u>0$, $w>0$, one
finds that the long-range stable fixed point is accessible only for
$\delta<\varepsilon<2\delta$, or $d<a<2+d/2$, a region where power
counting in Eq.~(\ref{h}) shows that the disorder is irrelevant.
In this sense the region III for the stability of the LR fixed point
is unphysical.
Formally, the first order results for $d<4$ read:
 \begin{equation}
 \label{9}
\nu = \left \{ \begin{array}{ll} \nu_{\rm pure}=
1/2 + \varepsilon/16,\; & \delta<\varepsilon/2, \\
\nu_{\rm LR}= 1/2 + \delta/8,\; &
\varepsilon/2 <\delta<\varepsilon.  \end{array}
\right.
\end{equation}
Thus, in this linear approximation  the
asymptotic behavior of polymers is governed by a distinct
exponent $\nu_{\rm LR}$ in the region III of the parameter plane $a,d$ .

\begin{figure} [!htb]
\epsffile{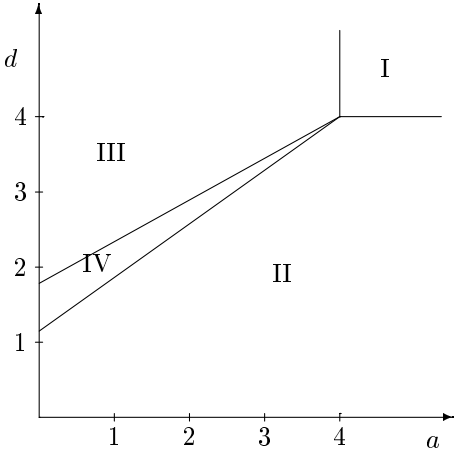}
\caption{\label{fig2}
The critical behavior of the $m=2$-magnet
in a medium with
long-range-correlated disorder in different
regions of the $d,a$-plane as predicted by the first order
$\varepsilon,\delta$--expansion.
Region I corresponds to the mean field behavior, in the region II the
critical exponents are the same as in the
medium with uncorrelated disorder, in  III the fixed point LR
is stable, in IV no accessible stable fixed points appear.
}
\end{figure}

Something similar happens if the $\varepsilon,\delta$--expansion is
applied to study models of $m$--vector magnets with
long-range-correlated quenched disorder. For comparison, using the
results of Weinrib and Halperin \cite{Weinrib} we get the phase
diagram presented for $m=2$ in the fig. \ref{fig2}.  Although the
critical behavior of the long-range-correlated universality class
appears there for $a<d$ where it is relevant by power counting
(region III in the fig.  \ref{fig2}) this region in the $d-a$ plane is
separated from the critical behavior of the $O(m=2)$ universality
class by a region, where no accessible stable fixed points are present
(IV in fig.  \ref{fig2}). It means that also in the case of magnets,
as well as for polymers the first order
$\varepsilon,\delta$--expansion leads to a controversial phase diagram
(compare figs. \ref{fig1} and \ref{fig2}).  So our first order
results should be considered as purely qualitative  and in order to
obtain a clear picture and more reliable information, we proceed to
higher order calculations.


\section{THE RESUMMATION AND THE RESULTS}\label{IV}
Fortunately, to investigate the 2-loop approximation in a fixed $d$ and $a$
approach we need not recalculate the intermediate expressions of
perturbation theory for the
vertex functions. Instead, we may make use of $m\to 0$ limit of the
appropriate $m$-vector model, investigated recently
\cite{Prudnikov}. Starting from the two-loop expressions of ref.
\cite{Prudnikov} for the RG functions of the $m$-vector magnet with
long-range-correlated disorder and applying the symmetry arguments
\cite{Kim,Blavats'ka01} for the polymer limit $m=0$ as explained in
Section \ref{II} we get the following expressions for the
$d=3$ RG functions of the model in Eq. (\ref{h}):
\begin{eqnarray}
\beta_u(u,w) & = & -u+u^2-(3f_1(a)-f_2(a))uw -
               \frac{95}{216}u^3+\frac{1}{8}b_2(a)u^2w- \nonumber \\
&&(b_3(a)+\frac{1}{4}b_6(a))uw^2 +f_3(a)w^2+b_5(a)w^3 \label{beta},
\end{eqnarray}
\begin{equation}\label{b2}
\beta_w(u,w)=-(4-a)w-(f_1(a)-f_2(a))w^2+\frac{uw}{2}+b_{10}(a)
w^3-\frac{23}{216}u^2w+ \frac{1}{4}b_{12}(a)uw^2
\end{equation}
\begin{equation}\label{b3}
\gamma_{\phi}(u,w)=\frac{1}{2}f_2(a)w+\frac{1}{108}u^2+c_1(a)w^2-
\frac{1}{4}c_2(a)uw,
\end{equation}
\begin{equation}\label{b4}
\gamma_{\phi^2}(u,w)=\frac{1}{4}u-\frac{1}{2}f_1(a)w-\frac{1}{16}u^2-c_3(a)w^2+
\frac{1}{4}c_4(a)uw.
\end{equation}
Here, the coefficients
$f_i(a)$ are expressed in terms of the one-loop integrals in Eq. (\ref{int}),
$b_i(a)$ and $c_i(a)$
originate from two-loop integrals and are tabulated in ref.
\cite{Prudnikov} for $d=3$ and different values of the parameter $a$
in the range $2\leq a \leq 3 $. The series are normalized by a
standard change of variables
$u\to\frac{3u}{4}I_1,w\to\frac{3w}{4}I_1$, so that the
coefficients of the terms $u,u^2$ in $ \beta_u $ become $1$ in
modulus.

The RG functions listed above have the form of a divergent series,
with zero radius of convergence \cite{Hardy}, familiar to the theory
of critical phenomena \cite{rgbooks}. If the nature of the divergence
is such that the series are asymptotic, then the situation is, at
least in principle, controllable: in this case a good estimate for
the sum of the series is obtained by keeping a certain number of the first
terms (``optimal truncation'') or applying an appropriate resummation
procedure.

For the case of the pure 3-dimensional $\phi^4$ theory it is known
that the perturbation series are asymptotic, and Borel-summability in
3 dimensions has been proven \cite{Eckmann}. The situation of the
random-site Ising model is less satisfactory than for the pure system
\cite{Folk00}.
For instance, the asymptotic parameter in the disordered system is
$\sqrt{\varepsilon}$ instead of $\varepsilon$, and the
$\beta$-functions, computed at two loops show no stable fixed points.
Bray et al \cite{Bray} and McKane \cite{McKane} studied the asymptotic
expansion for the free energy of the random-site Ising model in the
zero-dimensional case, and the model was found to be non-Borel
summable. However, recently \cite{Alvarez} the Borel summability of the
perturbation expansion for the zero-dimensional disordered Ising model
was proven analytically, provided that the summation is carried out in
two steps: first, in the coupling of the pure Ising model and subsequently
in the variance of the quenched disorder.

In our case, the summability of the series in Eq.~(\ref{beta}) is
open.  Nevertheless, we apply various kinds of resummation techniques
\cite{note1}, in order to obtain reliable quantitative results for the
problem under consideration and to check the stability of these
results.
\subsection{Chisholm-Borel resummation}
First, we employ a simple two-variable Chisholm-Borel resummation
technique \cite{Jug}. For our problem this turns out to be the most
effective one. The resummation procedure consists of several steps:

(i) starting from the initial RG function $f$ in the form of a
truncated series \cite{note1} in the variables
$u$ and $w$, one constructs its Borel image:
 $$f=\sum_{i,j}a_{ij}u^iw^j\rightarrow
\sum_{i,j}\frac{a_{ij}(ut)^i(wt)^j} {\Gamma(i+j+1)},$$ where
$\Gamma(x)$ is the Euler's gamma function;

(ii) the Borel image is extrapolated by a rational Chisholm
\cite{Chisholm,Baker}
approximant $[K/L](ut,wt)$ which is defined as the
ratio of two polynomials both in the variables $u$ and $w$ of degrees
$K$ and $L$ such that the truncated Taylor expansion of the
approximant is equal to that of the Borel image of the function $f$;

\begin{figure}[htbp]
\epsffile{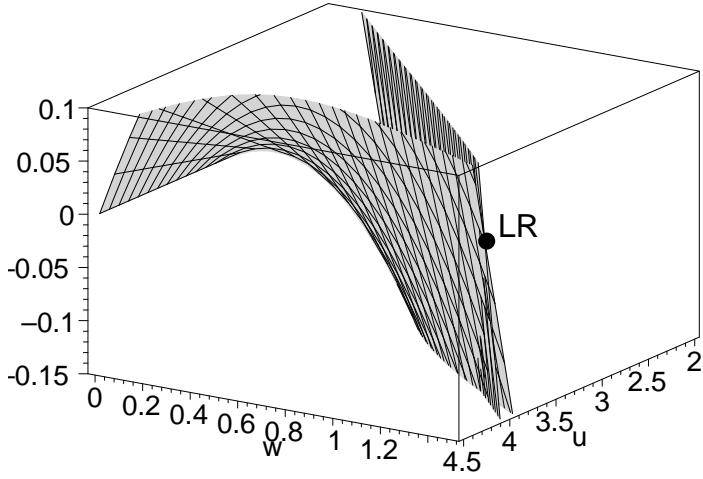}
\caption{ \label{fig3}
The Chisholm-Borel resummed 3d $\beta$-functions in the two-loop
approximation at $a=2.9$. The flat surface corresponds to the
$\beta_w$-function. The resummation restores the presence of a pure SAW
fixed point $(u^{\ast}=1.63, w^{\ast}=0)$ and leads to 
a new stable ``long-range-correlated'' fixed point. The coordinate
box is chosen to show the stable LR fixed point
$(u^{\ast}= 4.13, w^{\ast}=1.47)$
on the face of the box.}
\end{figure}

(iii) the resummed function $f^{\rm res}$ is then calculated as the
inverse Borel transform of this approximant:  $$
f^{\rm res}=\int\limits_0^{\infty}{\rm d}t \exp(-t)[K/L](ut,wt).  $$
There are a lot of possibilities to choose a Chisholm approximant in
two variables. The most natural way is to construct it such that, if
any of  $u$ or $w$ is equal to zero, it leads to the familiar results
for the reduced model. Here,
for the Borel-images of the $\beta$-functions Eqs. (\ref{beta}), (\ref{b2})
we have chosen the following approximants with linear denominators:
$$
\beta_u(u,w,t)^{\rm chis}=\frac{a_{1,0}ut+a_{2,0}u^2t^2+a_{1,1}
uwt^2+a_{0,2}w^2t^2
+a_{2,1}u^2wt^3+a_{1,2}uw^2t^3}{1+b_{1,0}ut+b_{0,1}wt},
$$
\begin{equation}
\beta_w(u,w,t)^{\rm chis}=\frac{c_{0,1}wt+
c_{0,2}w^2t^2+c_{2,1}u^2wt^3+
c_{1,2}uw^2t^3}{1+d_{1,0}ut+d_{0,1}wt}.
\label{sym}
\end{equation}
Note, that the polynomials in the numerators are chosen to
be symmetric in the variables $ u,w $. In fig.~\ref{fig3} we show the
resummed 3d $\beta$-functions in the $u,w$-plane for $a=2.9$.
In addition to the familiar fixed points describing
Gaussian chains and polymers we obtain the stable LR fixed point for
polymers in long-range-correlated disorder.
For comparison, we
depict the nonresummed functions in fig.~\ref{4}. Only the Gaussian
fixed point ($u^*=0,w^*=0$) is obtained without resummation.  In figure
\ref{fig5} we visualize the situation depicting the lines of zeroes of the
resummed $\beta$-functions at $a=2.9$ in the $u,w$-plane in the region
of interest.  The intersections of these curves correspond to the
fixed points. The corresponding values of the stable fixed point coordinates
and the stability matrix eigenvalues for different values of the correlation
parameter $a<3$ are given in our Table \ref{tab2}.

Substituting Eqs. (\ref{b3}), (\ref{b4})  into Eqs.(\ref{d1})--(\ref{d3})
we get the following expressions:
\begin{eqnarray}
\nu^{-1}(u,w)&=&
2-\frac{1}{4}u+\frac{(f_1(a)-f_2(a))}{2}w+\frac{(c_2(a)-c_4(a))}{4}uw+\frac
{23}{432}u^2+(c_3(a)-c_1(a))w^2,
\nonumber\\
\gamma (u,w)&=& 1+\frac{1}{8}u-\frac{f_1(a)}{4}w+\frac{(f_2(a)+4c_4(a)-
2f_1(a))}{32}uw- \frac{1}{64}u^2+ \nonumber \\
&&\frac{(f_1^2(a)-f_1(a)f_2(a)-8c_3(a))}{16}w^2.\label{exp}
\end{eqnarray}
This defines the critical exponents by $\nu^{-1}=\nu^{-1}(u^*,w^*)$ and
$\gamma=\gamma(u^*,w^*)$ at the stable accessible fixed point ($u^*,w^*$).
To calculate these exponents in the region where the LR fixed point
is stable, we again perform a resummation
of the series in Eq.~(\ref{exp}), using the following Chisholm
approximants:
\begin{equation}
\frac{c_{0,0}+c_{0,1}wt+c_{1,0}ut+c_{1,1}uwt^2}
{1+d_{1,0}ut+d_{0,1}wt}.
\end{equation}
The critical exponent $\eta$ is obtained from the scaling law in
Eq.~(\ref{d3}). The numerical values for $\nu$, $\gamma$ and $\eta$
are listed in  Table \ref{tab2} for $a=2.3, \dots ,2.9$.
Note, that for $a=3$, which
corresponds to short-range-correlated point-like defects, the
interactions $u$ and $w$ become of the same symmetry, so we pass
to one coupling $(u-w)$ and reproduce the well-known values of the critical
exponents for the pure SAW model. The numerical values corresponding to those
listed in table \ref{tab2} are in this case: $u^*= 1.63$, $\nu= 0.59$,
$\gamma=1.17$, $\eta= 0.02$,  $\omega= 0.64$.
As departing from the value $a=3$
downward to $2$ one notices a major increase of the value of the
coupling $u$, so the results are more reliable for $a$
close to 3. At some value $a=a_{\rm marg}$ the LR fixed point becomes
unstable.
This is explained by the following physical interpretation:
as noted in the introduction, the case $a=d-1$
(in our 3d approach $a=2$) corresponds to straight lines of
impurities of random orientation, and the absence of stable fixed
points for $a$ near $a=2$ suggests the
collapse of the polymer chain in such a medium.

It is difficult to estimate the accuracy of the numerical values
presented in Table \ref{tab2}. On one hand, it is the first
non-trivial result: in the one-loop approximation at fixed $d$, $a$
one does not encounter the LR fixed point so one can not estimate
deviations caused by different orders of the perturbation theory. On
the other hand, as is known from the experience with the studies of
magnets with long-range-correlated disorder \cite{Prudnikov}, the
convergence of the resummed series for the RG functions is worse than in
the pure (or the short-range-correlated) case. The resummed two-loop
expansions we exploited here give quite reliable estimates for the
exponents of SAWs on the pure ($d=3$) lattice (compare our two-loop
values $\nu=0.59$ and $\gamma=1.17$ with the most precise RG
estimates cited just after Eq. (\ref{nu})). In the case of
the short-range-correlated diluted magnets the comparison of recent
six-loop results \cite{Pelissetto} with the two-loop ones \cite{Jug}
brings about the accuracy of latter of the order of several percents.
While for our case no higher order calculations are available to
test the numerical accuracy of the data in Table \ref{tab2} the results clearly
confirm the presence of a new stable fixed point LR with critical
exponents that differ from those of the ``pure'' fixed point P.

In order to confirm the quantitative stability of the picture we obtained,
we have also used different non-symmetric approximants for $\beta_w$ instead
of the one given in Eq.~(\ref{sym}). As expected, this approach was less
effective in the sense that the region where a stable fixed point
could be established was reduced and the numerical values differed
from those given in Table \ref{tab2}. Nonetheless, the qualitative picture
is the same: an LR fixed point exists and is stable in some interval
$a_{\rm marg}\leq a < 3$.

\subsection{Subsequent resummation}
Secondly, we applied the method of
subsequent resummation, developed in the context of the $d=0$
dimensional diluted Ising model in ref. \cite{Alvarez} and
successfully used for the $d=3$ case in ref~\cite{Pelissetto}.  Here,
the summation was carried out first in the coupling $u$ and
subsequently in $w$.  Starting from the $\beta$-functions in
Eqs.~(\ref{beta}), (\ref{b2}) we rewrite them as series in the
variable $w$:
\begin{eqnarray*}
\beta_u(u,w)&=&-u+u^2-\frac{95}{216}u^3+w\left(u(f_2(a)-3f_1(a))+
\frac{b_2}{8}u^2\right)\\
&&+w^2\left(f_3(a)-u(b_3(a)+\frac{b_6(a)}{4})
\right)+b_5(a)w^3,
\end{eqnarray*}
\begin{eqnarray*}
\beta_w(u,w)&=&w\left(a-4+\frac{u}{2}-\frac{23}{216}u^2\right)+
 w^2\left(f_2(a)-f_1(a)+\frac{b_{12}(a)}{4}u\right)+b_{10}(a)w^3.
\end{eqnarray*}
We first perform a Pad\'e-Borel resummation of the coefficients at
different powers of $w$ in the variable $u$, where it is possible.
This results in a series  of the form:
\begin{equation}
f(u,w)=\sum_{i}A_i(u) w^i. \label{rr}
\end{equation}
The coefficients $A_i$ are some functions of $u$. Finally, the series
(\ref{rr}) are resummed in the variable $w$. While we do not expect
any high accuracy from this method, as far as the applicability has
not been proven for our problem, again the presence of a stable fixed
point LR for $a_{\rm marg} \leq a<d$ in this case confirms the
stability of a new type of critical behavior.

We note that in addition to the above procedures we have tried a
Pad\'e-Borel approximation for the summation of the RG functions.  To
treat the two variable case we used the representation in terms of a
resolvent series \cite{Watson,Baker} in a single auxiliary variable.
However, no fixed points with $w\neq0$ were found in the region of
interest. But note, that even for the weakly diluted quenched Ising
model this procedure does not lead to stable fixed points in the
three loop approximation \cite{Folk00}.
\subsection{Interpretation of the numerical results}

We may summarize and interpret our results as follows:

(i) A new stable fixed point (LR) for polymers in long-range-correlated
disorder is found for $d=3$, $a<d$, leading to critical exponents that
are different from those of the pure model;

(ii) There is a marginal value $a_{\rm marg}$ for the parameter $a$, below
which the stable fixed point is absent, indicating a chain collapse of the
polymer  for disorder that is stronger correlated.

(iii) The critical exponent $\nu$ increases with decreasing parameter
$a$, like in the Weinrib and Halperin case. But note, that the
relation $\nu=2/a$ does not hold. Physically this means that in weak
long-range-correlated disorder ($a>a_{\rm marg}$) the polymer coil swells with
increasing correlation of the disorder. The self avoiding path of the
polymer has to take larger deviations to avoid the defects of the
medium.

\section{CONCLUSIONS}\label{V}
In the present work, we have analysed the scaling behavior of
polymers in media with quenched defects that are correlated with a
correlation that decays like $\sim 1/x^a$ for large separations $x$.
This type of disorder is known to be relevant in magnetic systems
\cite{Weinrib,Prudnikov}, but the question about its relevance in the
polymer problem was so far not answered. To this end we applied
the field-theoretical RG approach, and performed renormalization for
fixed mass and zero external momenta \cite{Parisi}. In our study we
take special care of the symmetry properties of the effective
Hamiltonian of the system \cite{Kim}. Formerly \cite{Blavats'ka01} we
performed calculations up to the linear approximation, using a double
$\varepsilon, \delta$-expansion, as proposed for the magnetic problem
in the work of Weinrib and Halperin \cite{Weinrib}. While already this
study indicated the possibility of a new type of critical behavior in
such a system, it predicted such behavior for an unphysical range of
parameters.  A more sophisticated investigation at higher order of the
perturbation series was needed to confirm the existence of a distinct
polymer scaling behavior for long range correlated disorder.

We use 2-loop expressions for the RG functions that were recently obtained
for $m$-component systems in the fixed $d,a$
approach \cite{Prudnikov} and apply appropriate resummation techniques.
This way we confirm that in a medium with long-range-correlated
quenched disorder the swelling of the polymer coil is governed by a
distinct exponent $\nu_{\rm LR}$ that increases when the correlation
of the disorder is increased (i.e. $a$ is decreased).
When the correlation is too strong, i.e. $a$ is below some marginal
value $a_{\rm marg}>\approx 2$, then a crossover to the collapse of the polymer
is predicted.

\section{Appendix}
Here, we turn our attention to the 3d magnetic system. Recently
Ballesteros and Parisi \cite{Ballesteros} presented  Monte-Carlo
simulations of the site-diluted Ising model in three dimensions in the
presence of quenched disorder  with long-range-correlations. The
values of the corrections-to-scaling exponents are of great interest in
the interpretation of such simulations. In previous work, dedicated to
3d magnets with long-range-correlated disorder
\cite{Prudnikov,Dorogovtsev,Cardy} these exponents have
not been calculated. Here we carry out these calculations based on the
$\beta$-functions of the model Eq. (\ref{4}) in the 2-loop approximation,
as presented in ref.~\cite{Prudnikov}.

The correction-to-scaling exponent is defined as the minimal
stability matrix eigenvalue in the stable accessible fixed point.  We carry
out the investigation in new variables
$(u,v,w) \to (u,v,v+w)$, as proposed by Dorogovtsev
\cite{Dorogovtsev} and perform the Pad\'e-Borel resummation of
the $\beta$-functions. The results are presented in Table
\ref{tab6}.


\clearpage

 \begin{table}[htp]
 \begin{center}
\begin{tabular}{|c|cccc|}
\hline
Fixed Point & $u^{\ast}$ & $w^{\ast}$ & $\omega_1$ &
$\omega_2$\\ \hline Gaussian (G) & $0$ & $0$ & $-\varepsilon$ &
$-\delta$\\ \hline Pure SAW (P) & $\varepsilon$ & $0$ &
$\varepsilon$ & $\varepsilon/2-\delta$\\ \hline Long-range (LR)  &
$\frac{2\delta^2}{(\varepsilon-\delta)}$  &
 $-\frac{\delta(\varepsilon-2\delta)}{(\varepsilon-\delta)}$ &
$\phantom{555555}\frac{1}{2}\{\varepsilon-4\delta\pm\sqrt
{\varepsilon^2-4\varepsilon\delta+8\delta^2}\}$ &  \\
\hline
\end{tabular}
\caption{\label{tab1}Fixed points and stability matrix eigenvalues in the first order
of the
$\varepsilon, \delta$~-~expansion.}
\end{center}
\end{table}

\begin{table}[htp]
\begin{center}
\begin{tabular}{|c|c|c|c|c|c|c|}
\hline
$a$ &  $u^*$  &  $w^*$ &  $\nu$ & $\gamma$ & $\eta$ & $\omega_{1,2}$ \\
 \hline
2.9 & 4.13 & 1.47 & 0.64 & 1.25 & 0.04 &    0.25 $\pm$ 0.62 i \\
2.8 & 4.73 & 1.68 & 0.64 & 1.26 & 0.04 &    0.22 $\pm$ 0.76 i \\
2.7 & 5.31 & 1.81 & 0.65 & 1.28 & 0.03 &    0.18 $\pm$ 0.89 i \\
2.6 & 5.89 & 1.87 & 0.66 & 1.29 & 0.03 &    0.15 $\pm$ 0.99 i \\
2.5 & 6.48 & 1.89 & 0.66 & 1.31 & 0.02 &    0.11 $\pm$ 1.09 i \\
2.4 & 7.10 & 1.87 & 0.67 & 1.33 & 0.01 &    0.07 $\pm$ 1.18 i \\
2.3 & 7.76 & 1.84 & 0.68 & 1.36 & 0.01 &    0.03 $\pm$ 1.26 i \\
\hline
\end{tabular}
\caption { \label{tab2} Stable fixed point of the  3d two-loop
$\beta$-functions, resummed by the Chisholm-Borel method, the
corresponding critical exponents and the stability matrix
eigenvalues at various values of $a$.}
\end{center} \end{table}

\begin{table}[!htp]
\begin{center}
\begin{tabular}{|c|cccccccccc|}
\hline
 $a$  & 2 & 2.1  & 2.2 & 2.3 & 2.4 & 2.5 & 2.6 & 2.7 & 2.8 & 2.9 \\
\hline
$\omega(m=1)$ & 0.80& 0.81  & 0.83  & 0.87 & 0.94  & 1.14 &
 1.07  &0.87 & 0.71  & 0.69  \\
 \hline
$\omega(m=2)$ & 1.15  & 1.08  &0.93  &0.86 & 0.81 & 0.68  &
0.59  & 0.57  & 0.55 & 0.54 \\
\hline
 $\omega(m=3)$ &  0.88 & 0.83& 0.76 & 0.67 & 0.62 & 0.61 &
0.60 & 0.60 &0.59 & 0.68   \\
\hline
\end{tabular}
\caption
{ \label{tab6} The correction-to-scaling exponents $\omega$ for the phase
transition in the 3d $m$-vector  model with $m=1,2,3$.}
\end{center}
\end{table}

\begin{figure}[htbp]
\epsffile{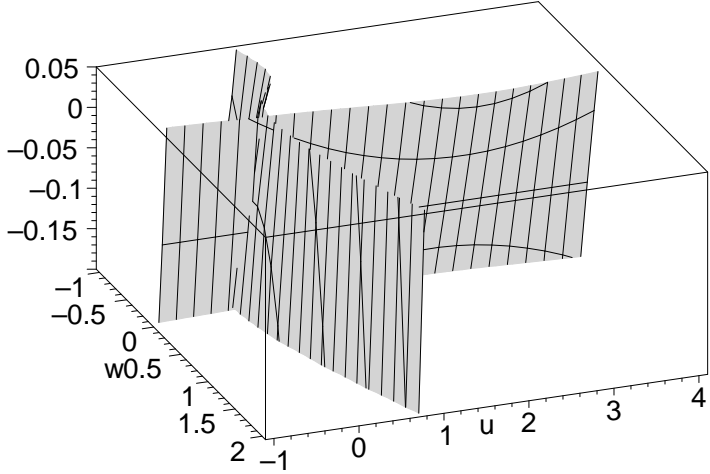}
\caption{ \label{fig4}
The non-resummed 3d $\beta$-functions at $a=2.9$. The intersection of
the $\beta_u$ and $\beta_w$ surfaces with the $u-w$ plane give the fixed
points. Only the Gaussian  fixed point $u^{\ast}=w^{\ast}=0$ is present
without resummation.}
\end{figure}

\begin{figure}[htbp]
\begin{centering}
\setlength{\unitlength}{0.4mm}
\begin{picture}(150,150)
\epsfxsize=20mm
\epsfysize=20mm
\put(-30,-10){\epsffile[1 3 200 200]{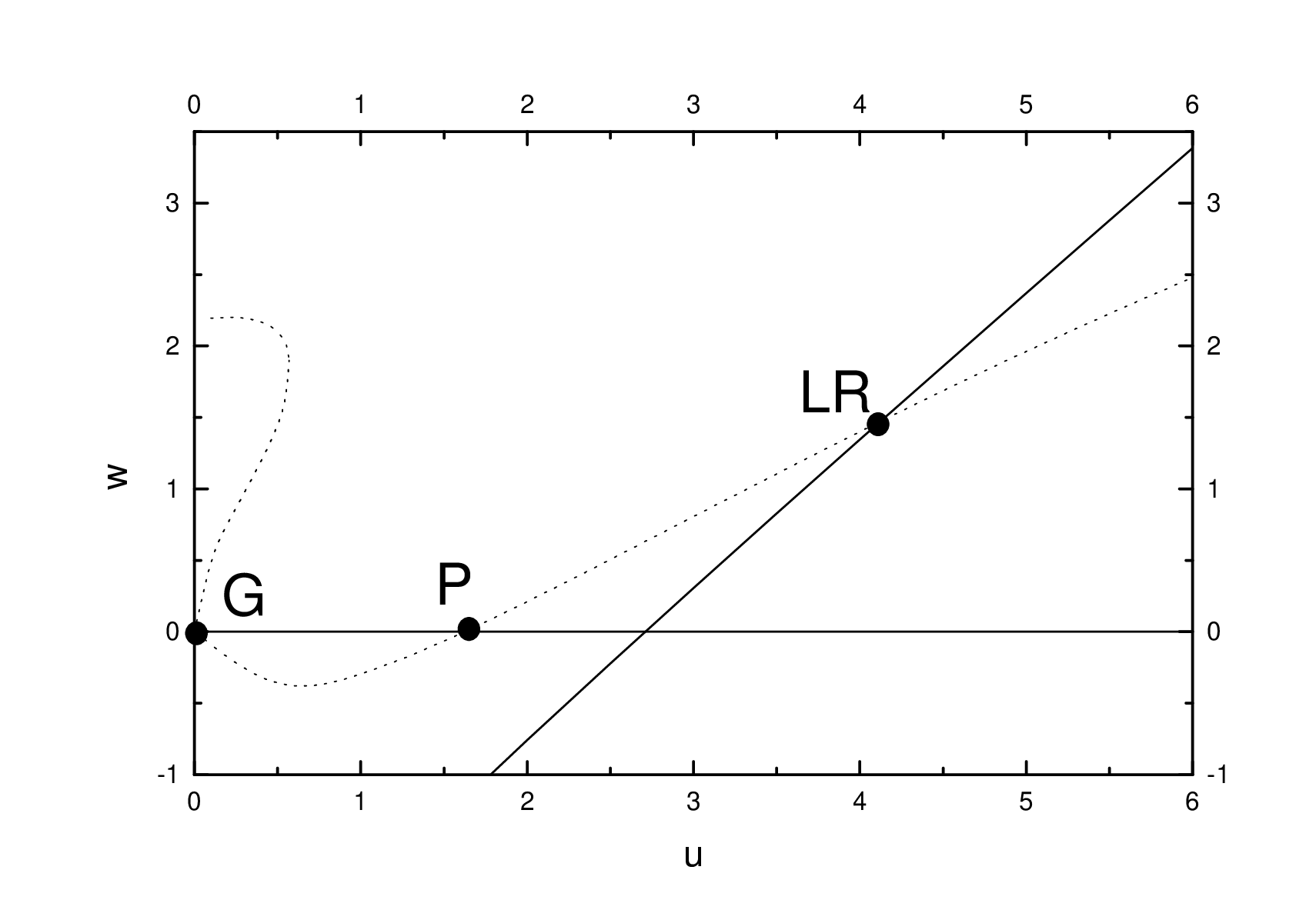}}
\end{picture}\\
\end{centering}
\caption{ \label{fig5}
The lines of zeroes of the 3d $\beta$-functions (\ref{beta}), (\ref{b2})
resummed by the Chisholm-Borel method at $a=2.9$. The dashed line
corresponds to $\beta_u=0$, the solid lines depict $\beta_w=0$.
The intersections of the dashed and solid lines give three fixed points
shown by filled circles at
$u^{\ast}=0, w^{\ast}=0$ (G), $u^{\ast}=1.63, w^{\ast}=0$ (P), and
$u^{\ast}=4.13, w^{\ast}=1.47$ (LR). The fixed point LR is stable.}
\end{figure}

\begin{thebibliography}{22}
\bibitem{note1}
We perform our analysis of the two-loop expressions for the RG functions
in the fixed $d,a$ technique. It is known that for other models with
disorder, the $\varepsilon$-expansion leads to less reliable results
\cite{Folk00}.
\end{thebibliography}
\end{document}